\documentclass[onecolumn,english]{revtex4}

\usepackage{natbib,hyperref}

\usepackage[T1]{fontenc}
\usepackage[latin9]{inputenc}
\usepackage{color}
\usepackage{graphicx}
\usepackage{amssymb,amsmath}
\usepackage{babel}

\def\be{\begin{equation}}
\def\ee{\end{equation}}

\makeatletter

\@ifundefined{definecolor}
{\usepackage{color}}{}

\newcommand{\beq}{\begin{equation}}
\newcommand{\eeq}{\end{equation}}  
\newcommand{\ba}{\begin{eqnarray}}
\newcommand{\ea}{\end{eqnarray}}
\newcommand{\bef}{\begin{figure}}
	\newcommand{\eef}{\end{figure}}

\def\x{{\bf x}}

\begin{document}

\title{Temperature dependence of COVID-19 transmission}


	\author{Alessio Notari$^{1}$}
	\email{notari@fqa.ub.edu}
	
	\affiliation{$^{1}$ Departament de F\'isica Qu\`antica i Astrofis\'ica \& Institut de Ci\`encies del Cosmos (ICCUB), Universitat de Barcelona, Mart\'i i Franqu\`es 1, 08028 Barcelona, Spain}

\begin{abstract}
\begin{center}
\textbf{Abstract}
\end{center}
The recent coronavirus pandemic follows in its early stages an almost exponential expansion, with the number of cases as a function of time reasonably well fit by $N(t)\propto e^{\alpha t}$, in many countries. We analyze the rate $\alpha$ in different countries, choosing as a starting point in each country a threshold of 30 total cases and fitting for the following 12 days, capturing thus the early exponential growth in a rather homogeneous way. We look for a link between  the rate $\alpha$ and the average temperature  $T$ of each country, in the month of the epidemic growth. We analyze a {\it base} set of 42 countries, which developed the epidemic at an earlier stage, an {\it intermediate} set of 88 countries and an {\it extended} set of 125 countries, which developed the epidemic more recently.
Fitting with a linear behavior $\alpha(T)$, we find increasing evidence in the three datasets for a decreasing growth rate as a function of $T$, at $99.66\%$C.L., $99.86\%$C.L. and $99.99995 \%$ C.L. ($p$-value $5 \cdot 10^{-7}$, or 5$\sigma$ detection) in the {\it base}, {\it intermediate}  and {\it extended} dataset, respectively.  The doubling time is expected to increase by $40\%\sim 50\%$, going from $5^\circ$ C to $25^\circ$ C.
In the {\it base} set, going beyond a linear model, a  peak at about $(7.7\pm 3.6)^\circ C$ seems to be present in the data, but such evidence disappears for the larger datasets. 
Moreover we have analyzed the possible existence of a bias: poor countries, typically located in warm regions, might have less intense testing. By excluding countries below a given GDP per capita from the dataset, we find that this affects our conclusions only slightly and only for the {\it extended} dataset. The significance always remains high, with a $p$-value of about $10^{-3}-10^{-4}$ or less. Our findings give hope that, for northern hemisphere countries, the growth rate should significantly decrease as a result of both warmer weather and lockdown policies. In general the propagation should be hopefully stopped by strong lockdown, testing and tracking policies, before the arrival of the next cold season.

\end{abstract}

\maketitle

\section{ Introduction}

The recent coronavirus (COVID-19) pandemic is having a major effect in many countries, which needs to be faced with the highest degree of scrutiny. An important piece of information is whether the growth rate of the confirmed cases among the population could decrease with increasing temperature.  Experimental research on related viruses found indeed a decrease at high temperature and humidity~\cite{Chan:2011}. We try to address this question using available epidemiological data. A similar analysis for the data from January 20 to February 4, 2020, among 403 different Chinese cities, was performed in~\cite{Wang2020.02.22.20025791} and similar studies were recently performed in~\cite{Araujo2020.03.12.20034728,Bukhari2020,Wang2020,Sajadi2020,Luo2020.02.12.20022467}. 
The paper is organized as follows. In section~\ref{methods} we explain our methods, in section~\ref{results} we show the results of our analysis and in section~\ref{conclusions} we draw our conclusions.

\section{Method} \label{methods}
We start our analysis from the empirical observation that the data for the coronavirus disease in many different countries follow a common pattern: once the number of confirmed cases reaches order 10 there is a very rapid subsequent growth, which is well fit by an exponential behavior. The latter  is typically a  good approximation for the following couple of weeks and, after this stage of {\it free} propagation, the exponential growth typically gradually slows down, probably due to other effects, such as: lockdown policies from governments, a higher degree of awareness in the population or the tracking and isolation of the positive cases. 

Our aim is to see whether the temperature of the environment has an effect, and for this purpose we choose to analyze the first stage of {\it free} propagation in a selected sample of countries. We choose our sample using the following rules:
\begin{itemize} 
\item we start analyzing data from the first day in which the number of cases in a given country reaches a reference number $N_i$, which we choose to be $N_i=30$~\footnote{In practice we choose, as the first day, the one in which the number of cases $N_i$ is closest to 30. In some countries, such a number $N_i$ is repeated for several days; in such cases we choose the last of such days as the starting point. For the particular case of China, we started from January 16th, with 59 cases, since the number before that day was essentially frozen.};
\item we include only countries with at least 12 days of data, after this starting point.
\end{itemize}
The data were collected from~\footnote{https://www.ecdc.europa.eu/en/geographical-distribution-2019-ncov-cases}. We then fit the data for each country with a simple exponential curve $N(t)=N_0 \, e^{\alpha t}$, with 2 parameters, $N_0$ and $\alpha$; here $t$ is in units of days. In the fit we used Poissonian errors, given by $\sqrt{N}$, on the daily counting of cases. We associated then to each country an average temperature $T$, for the relevant weeks, which we took from~\footnote{ https://en.wikipedia.org/wiki/List\_of\_cities\_by\_average\_temperature}. More precisely: if for a given country the average $T$ is tabulated only for its capital city, we directly used such a  value. If, instead, more cities are present for a given country, we used an average of the temperatures of the main cities, weighted by their population~\footnote{The only two exceptions to this procedure are: Japan and U.S.A.. For Japan we have subdivided into three regions: Hokkaido, Okinawa and the rest of the country, using respectively the temperatures of Sapporo, Naha and Tokio. For the U.S.A. we  used the national average of about 5.3 degrees from https://www.ncdc.noaa.gov/sotc/national/201903. For Ecuador, we used the average $T=27.5^\circ C$ of Guayaquil, the main site for the disease. For Brunei we used $T=27 \,^\circ C$ and for Morocco we used the average Temperature for Casablanca, $T=14.7^\circ C$, from https://en.climate-data.org. For Jordan we used $T=17^\circ C$, from https://www.weather-atlas.com/en/jordan/amman-weather-march.}. For most countries we used the average temperature for the month of March, with a few exceptions~\footnote{For China, South Korea, Singapore, Iran, Taiwan and Japan we considered an interpolating function of the temperature for the months of January, February and March and we took an average of such function in the relevant 12 days of the epidemic.}.

We analyzed three datasets. A first list of countries was selected on March 26th.
The list of such {\it base} dataset includes 42 countries: Argentina, Australia, Belgium, Brazil, Canada, Chile, China, Czech Republic, Denmark, Egypt, Finland, France, Germany, Greece, Iceland, India, Indonesia, Iran, Ireland, Israel, Italy, Lebanon, Japan, Malaysia, Netherlands, Norway, Philippines, Poland, Portugal, Romania, Saudi Arabia, Singapore, Slovenia, South Korea, Spain, Sweden, Switzerland, Taiwan, Thailand, United Arab Emirates, United Kingdom, U.S.A..

An additional set of countries was added to the first dataset on April 1st, reaching a total of 88 countries. The added countries, in this {\it intermediate} set, are: Albania, Andorra, Algeria, Armenia, Austria, Bahrain, Bosnia and Herzegovina, Brunei, Bulgaria, Burkina Faso, Cambodia, Colombia, Costa Rica, Croatia, Cyprus, Dominican Republic, Ecuador, Estonia, Hungary, Iraq, Jordan, Kazakhstan, Kuwait, Latvia, Lithuania, Luxembourg, Malta, Mexico, Moldova, Morocco, New Zealand, North Macedonia, Oman, Panama, Pakistan, Peru, Qatar, Russia, Senegal, Serbia, Slovakia, South Africa, Tunisia, Turkey, Ukraine, Uruguay, Vietnam.

Finally an {\it extended} set has been studied on April 14th~\footnote{Only countries with at least 300.000 inhabitants have been considered in this dataset.}, adding the following countries to the previous dataset: Belarus, Bolivia, Cameroon, 
Congo, 
 Cote d'Ivoire,  Cuba, 
  Democratic Republic of Congo, Djibouti,  
 El Salvador, Georgia,  Ghana, 
 Guatemala, Guinea, Honduras, Jamaica,  Kenya, Kosovo,  Kyrgyzstan, Madagascar, 
Mali, Mauritius, Montenegro, Niger, Nigeria, Paraguay, 
 Puerto Rico, Rwanda, 
  Sri Lanka,  Togo, 
 Trinidad and Tobago ,  Uganda, Uzbekistan, Venezuela, Zambia.

Using such datasets for $\alpha$ and $T$ for each country, we fit with two functions $\alpha(T)$, as explained in the next section. Note that the statistical errors on the $\alpha$ parameters, considering Poissonian errors on the daily counting of cases, are typically much smaller than the spread of the values of $\alpha$ among the various countries. This is due to systematic effects, which are dominant, as we will discuss later on. For this reason we disregarded statistical errors on $\alpha$. The analysis was done using  the software {\it Mathematica}, from Wolfram Research, Inc..

\section{Results} \label{results}

\begin{figure}
\begin{center}
\vspace*{3mm}
\includegraphics[scale=0.6]{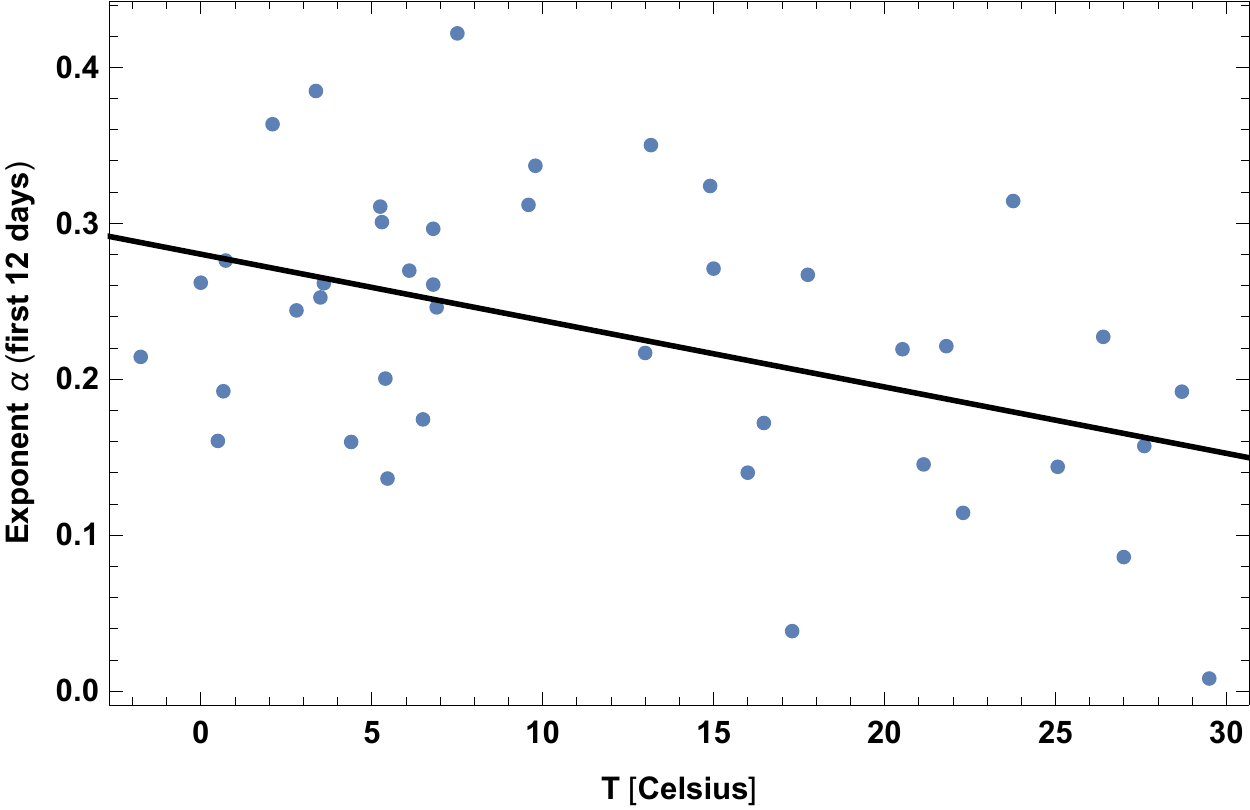} 
\caption{Exponent $\alpha$ for each country vs.~average temperature $T$, for the relevant period of time, as defined in the text,  for the base set of 42 countries. We show the data points and the best-fit for the linear interpolation. \label{fig1}}
\end{center}
\end{figure}

We first fit the {\it base} dataset, with a simple linear function $\alpha(T)=\alpha_0+\beta \, T$, to look for an overall decreasing behavior. 
Results for the best fit, together with our data points, are shown in fig.~\ref{fig1}. The estimate, standard deviation, confidence intervals for the parameters, together with the significance and the explained variance, $R^2$, are shown in Table~\ref{tab1}.
From such results a clear decreasing trend is visible, and indeed the slope $\beta$ is negative, at $99.66\%$ C.L. ($p$-value 0.0034).

\begin{table}[!htb]

\hspace*{4em}
\begin{tabular}{cc}
    \begin{minipage}{.5\linewidth}
       \begin{tabular}{|l|c|c|c|c|}
 \hline

Parameter & estimate & $\sigma$ & 95\% lower & 95\% upper \\ \hline
$\alpha_0$ &$0.280$ & $0.021$ & $0.238$ & $0.321$ \\
$\beta$ &$-0.00425$ & $0.00136$ & $-0.00701$ & $-0.00149$ \\
\hline
 \end{tabular} 
    \end{minipage} &
\hspace*{-4em}
    \begin{minipage}{.5\linewidth}
             \begin{tabular}{|l|c|c|c|c|}
 \hline
$R^2$ &  $0.196$ \\ \hline
 $p$-value & $0.0033$     \\  \hline
 \end{tabular}
    \end{minipage} 
\end{tabular}

    \caption{In the left panel: best-estimate, standard deviation ($\sigma$) and $95\%$ C.L. intervals for the parameters of the linear interpolation,  for the {\it base} set of 42 countries. In the right panel: $R^2$ for the  best-estimate and $p-$value of a non-zero $\beta$.}
    \label{tab1}

\end{table}

%
%
%

However, the linear fit is able to explain only a small part of the variance of the data, with $R^2=0.196$, and its adjusted value $R^2_{\rm adjusted}=0.175$, clearly due to the presence of many more factors.

\begin{figure}
\begin{center}
\vspace*{3mm}
\includegraphics[scale=0.6]{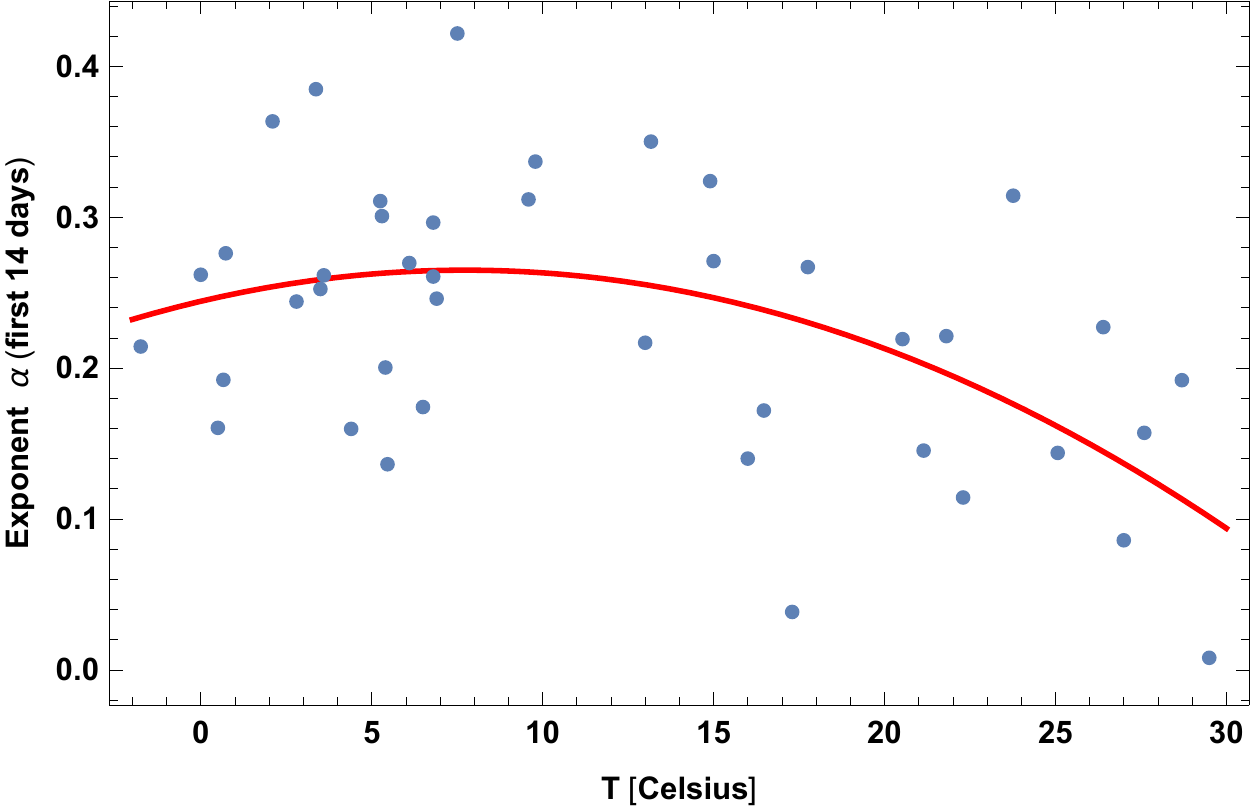}
\caption{Exponent $\alpha$ for each country vs. average temperature $T$, as defined in the text, for the base set of 42 countries. We show here the quadratic best-fit. \label{fig2}}
\end{center}
\end{figure}

In addition, a decreasing trend is also visible in this dataset, below about 10$^\circ C$. For this reason we also fit with a quadratic function $\alpha(T)=\alpha_0-\beta (T-T_M)^2 \, $.
Results for the quadratic best fit are presented in fig.~\ref{fig2} and in Table~\ref{tab2}.
From such results a peak is visible at around $T_M\approx 8^\circ C$. The quadratic model is able to explain a  slightly larger part of the variance of the data, since $R^2 \approx 0.27$~\footnote{Here $R^2$ is defined as $R^2\equiv 1-\frac{SS_R}{SS_T}$, where $SS_R$ is the  residual sum of squares and $SS_T$ is the sum of the squared differences between the $\alpha$ values and their mean value.}. Moreover, despite the presence of an extra parameter, one may quantify the improvement of the fit, using for instance the  Akaike Information Criterion (AIC) for model comparison,  $\Delta {\rm AIC}\equiv 2 \Delta k-2 \Delta \ln ({\cal L})$, where $\Delta k$ is the increase in the number of parameters, compared to the simple linear model, and $\Delta \ln ({\cal L})$ is the change in the maximum log-likelihood between the two models. This gives $\Delta {\rm AIC}=-2.1$, slightly in favor of the quadratic model.

\begin{table}[!htb]

\hspace*{4em}
\begin{tabular}{cc}
    \begin{minipage}{.5\linewidth}
       \begin{tabular}{|l|c|c|c|c|}
 \hline

Parameter & estimate & $\sigma$ & 95\% lower & 95\% upper \\ \hline
$\alpha_0$ &$0.264$ & $0.0159$ & $0.2325$ & $ 0.2972$ \\
$\beta$ &$0.000345$ & $0.000173$ & $-5.104\cdot10^{-6}$ & $0.000694$ \\
$T_M$ &$7.73$ & $3.64$ & $0.37$ & $15.1$ \\
\hline
 \end{tabular} 
    \end{minipage} &
\hspace*{-4em}
    \begin{minipage}{.5\linewidth}
             \begin{tabular}{|l|c|c|c|c|}
 \hline
$R^2$ &  $0.27$ \\ \hline
$p$-value ($\beta$) & $0.053$     \\  \hline
 \end{tabular}
    \end{minipage} 
\end{tabular}

    \caption{In the left panel: best-estimate, standard deviation ($\sigma$) and $95\%$ C.L. intervals for the parameters of the quadratic interpolation,  for the {\it base} set of 42 countries. In the right panel: $R^2$ for the  best-estimate and $p-$value of a non-zero $\beta$.}
    \label{tab2}
\end{table}

%
%
%
%
%
%
%
%
%
%

\begin{figure}
\begin{center}
\vspace*{3mm}
\includegraphics[scale=0.6]{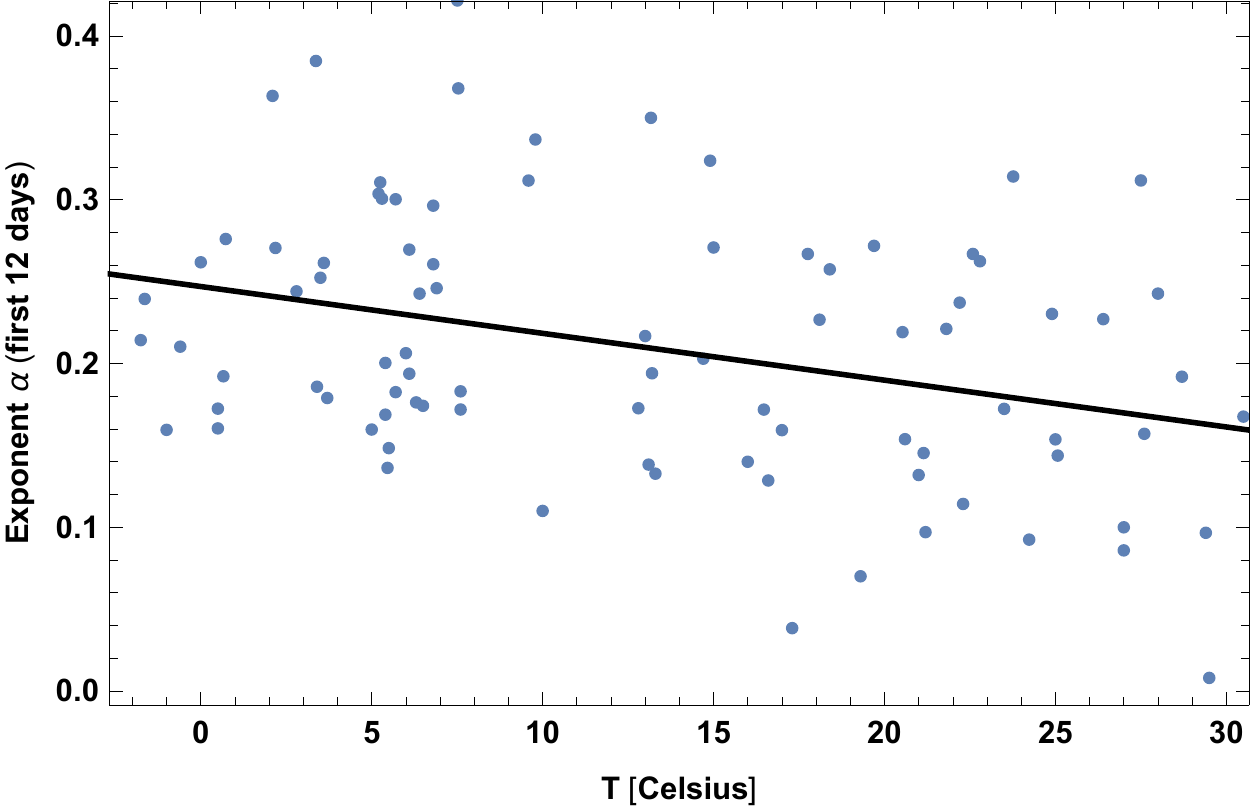} 
\caption{Exponent $\alpha$ for each country vs. average temperature $T$, for the relevant period of time, as defined in the text, for the {\it intermediate} set of 88 countries. We show the data points and the best-fit for the linear interpolation. \label{fig3}}
\end{center}
\end{figure}

We repeated then the same analysis for the {\it intermediate} dataset of 88 countries and for the {\it extended} dataset of 125 countries.
Results for the linear fit of the {\it intermediate} sample are shown in fig.~\ref{fig3} and in Table~\ref{tab3}.
The slope $\beta$ is smaller in absolute value, but the significance actually slightly increases, since a zero slope is excluded at $99.86\%$ C.L. ($p$-value 0.0014). Now $R^2=0.11$ and $R^2_{\rm adjusted}=0.10$.

In this sample the quadratic trend is not visible anymore, and indeed the AIC does not prefer the quadratic fit: $\Delta {\rm AIC}=+0.9$ compared to the linear fit, in disfavor of the quadratic model. The $R^2$ is also practically the same as in the linear fit.

\begin{table}[!htb]

\hspace*{4em}
\begin{tabular}{cc}
    \begin{minipage}{.5\linewidth}
       \begin{tabular}{|l|c|c|c|c|}
 \hline

Parameter & estimate & $\sigma$ & 95\% lower & 95\% upper \\ \hline
$\alpha_0$ &$0.247$ & $0.0138$ & $0.220$ & $0.275$ \\
$\beta$ &$-0.00286$ & $0.000867$ & $-0.00458$ & $-0.00113$ \\
\hline
 \end{tabular} 
    \end{minipage} &
\hspace*{-4em}
    \begin{minipage}{.5\linewidth}
             \begin{tabular}{|l|c|c|c|c|}
 \hline
$R^2$ &  $0.11$ \\ \hline
 $p$-value & $0.0014$     \\  \hline
 \end{tabular}
    \end{minipage} 
\end{tabular}

    \caption{In the left panel: best-estimate, standard deviation ($\sigma$) and $95\%$ C.L. intervals for the parameters of the linear interpolation,  for the {\it intermediate} set of 88 countries. In the right panel: $R^2$ for the  best-estimate and $p-$value of a non-zero $\beta$.}
    \label{tab3}
\end{table}

\begin{figure}
\begin{center}
\vspace*{3mm}
\includegraphics[scale=0.6]{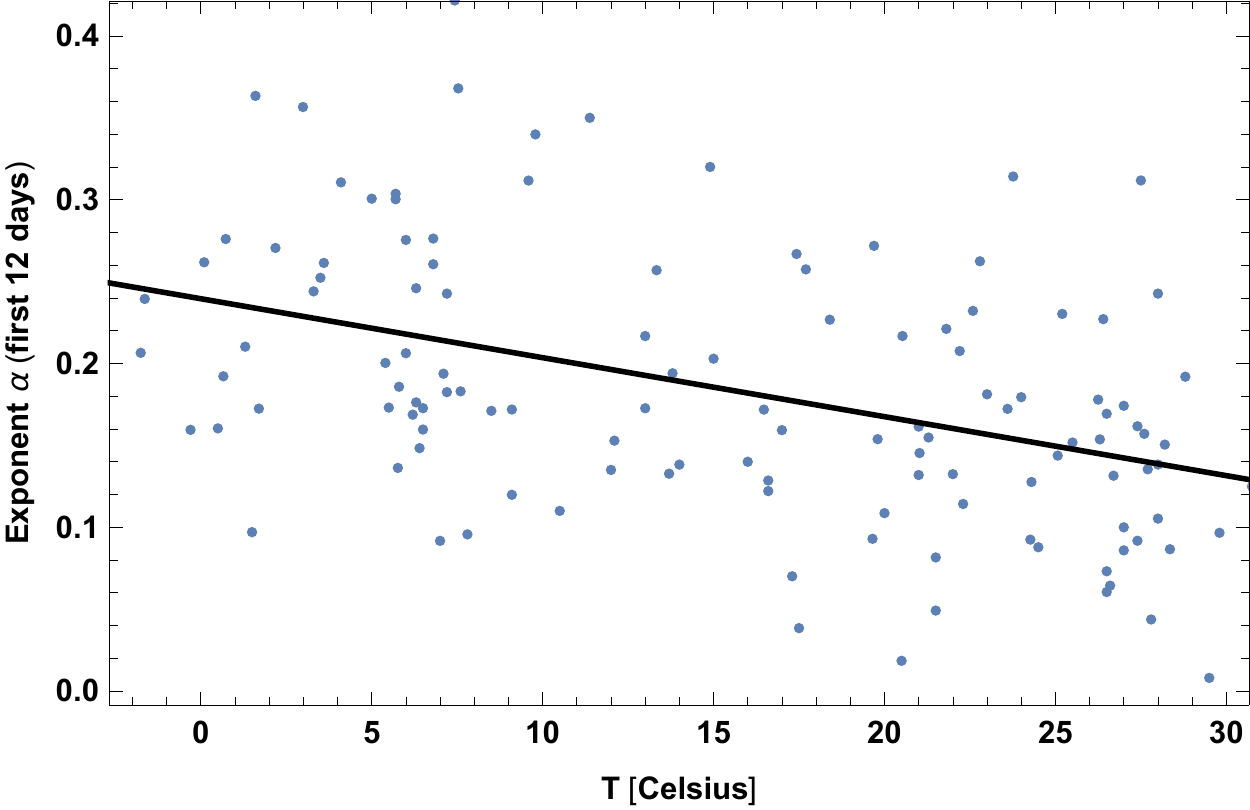} 
\caption{Exponent $\alpha$ for each country vs. average temperature $T$, for the relevant period of time, as defined in the text, for the {\it extended} set of 125 countries. We show the data points and the best-fit for the linear interpolation. \label{fig4}}
\end{center}
\end{figure}

For the {\it extended} sample results of the linear fit are shown in fig.~\ref{fig4} and in Table~\ref{tab4}.
The slope $\beta$ becomes larger and, most importantly, the significance highly increases, since a zero slope is now excluded at   $99.99995 \%$ C.L. ($p$-value $5 \cdot 10^{-7}$, or 5$\sigma$ detection, translated in the language of a Gaussian distribution). Now $R^2=0.19$ and $R^2_{\rm adjusted}=0.18$.

In this dataset, which extends to April 14th, a few anomalies are however present: in the case of Bangladesh and Thailand it is possible to see that the exponential growth became much faster after the initial 12 days. We have checked what happens by using a different interval of time for these 2 cases, instead of the standard 12 days. Namely we have used 44 days for Thailand and 21 days for Bangladesh, which give the maximal value of $\alpha$ in both cases. The results for the linear fits using such corrected values is shown in Table~\ref{tab5}. The significance is lower, but still very high: $p$-value $4.6 \cdot 10^{-6}$, or 4.6$\sigma$ detection, translated in the language of a Gaussian distribution.
%

\begin{table}[!htb]

\hspace*{4em}
\begin{tabular}{cc}
    \begin{minipage}{.5\linewidth}
       \begin{tabular}{|l|c|c|c|c|}
 \hline

Parameter & estimate & $\sigma$ & 95\% lower & 95\% upper \\ \hline
$\alpha_0$ & $0.2396 $ & $ 0.01251 $ & $  0.2148 $ & $  0.2643$
\\
$\beta$ & $ -0.003602$ & $  0.0006782 $ & $ -0.004944 $ & $ -0.002258 $
 \\
\hline
 \end{tabular} 
    \end{minipage} &
\hspace*{-4em}
    \begin{minipage}{.5\linewidth}
             \begin{tabular}{|l|c|c|c|c|}
 \hline
$R^2$ &  $0.19$ \\ \hline
 $p$-value & $5\cdot 10^{-7}$     \\  \hline
 \end{tabular}
    \end{minipage} 
\end{tabular}

 \caption{In the left panel: best-estimate, standard deviation ($\sigma$) and $95\%$ C.L. intervals for the parameters of the linear interpolation, for the {\it extended} set of 125 countries. In the right panel: $R^2$ for the  best-estimate and $p-$value of a non-zero $\beta$.}
 \label{tab4}
\end{table}

\begin{table}[!htb]

\hspace*{4em}
\begin{tabular}{cc}
    \begin{minipage}{.5\linewidth}
       \begin{tabular}{|l|c|c|c|c|}
 \hline

Parameter & estimate & $\sigma$ & 95\% lower & 95\% upper \\ \hline
$\alpha_0$ & $ 0.2364 $ & $ 0.01235 $ & $  0.2120 $ & $ 0.2609 $ \\
$\beta$ & $-0.00321 $ & $ 0.0006699 $ & $ -0.004538 $ & $ -0.001885 $ \\
\hline
 \end{tabular} 
    \end{minipage} &
\hspace*{-4em}
    \begin{minipage}{.5\linewidth}
             \begin{tabular}{|l|c|c|c|c|}
 \hline
$R^2$ &  $0.16$ \\ \hline
 $p$-value & $4.6\cdot 10^{-6}$     \\  \hline
 \end{tabular}
    \end{minipage} 
\end{tabular}

 \caption{In the left panel: best-estimate, standard deviation ($\sigma$) and $95\%$ C.L. intervals for the parameters of the linear interpolation, for the {\it extended} set of 125 countries. Here Thailand and Bangladesh have been corrected for, as explained in the text. In the right panel: $R^2$ for the  best-estimate and $p-$value of a non-zero $\beta$.}
 \label{tab5}
\end{table}

%
%
%
%

\begin{figure}
\begin{center}
\vspace*{3mm}
\includegraphics[scale=0.65]{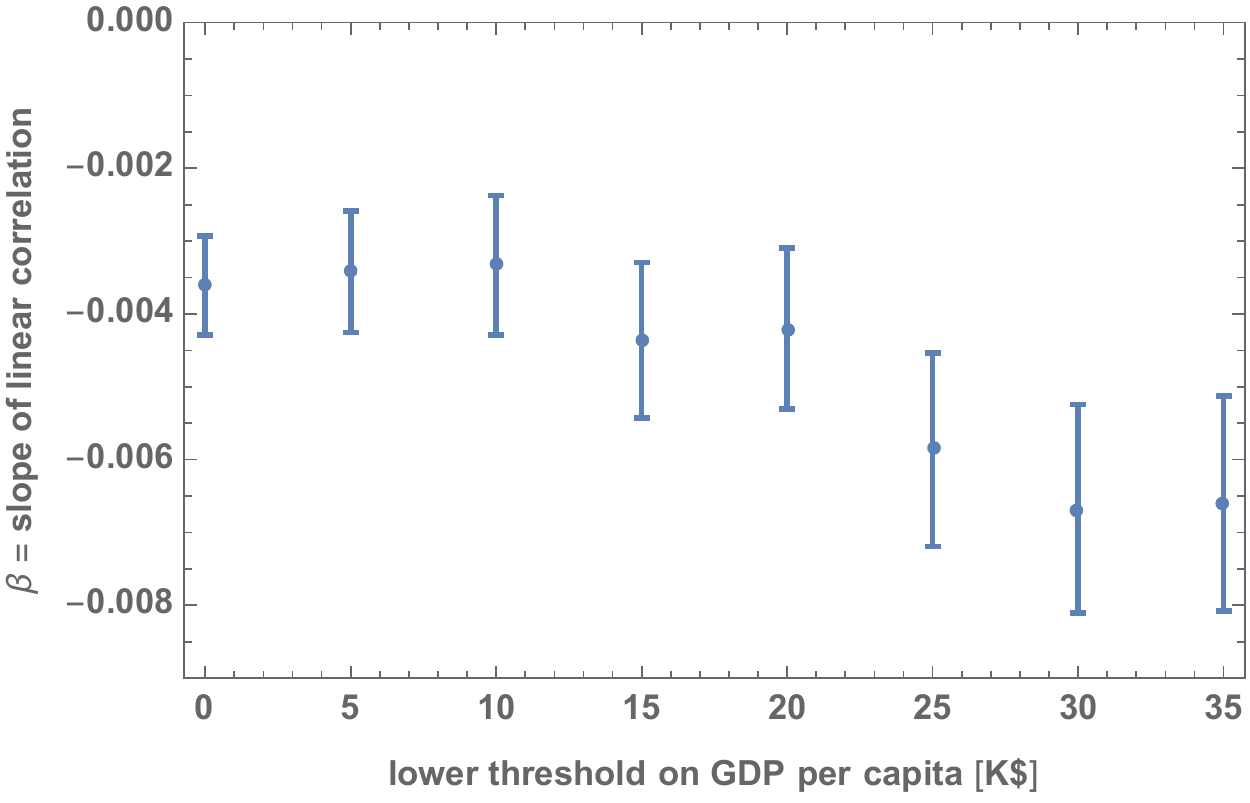} 
\caption{We show the best estimate and the standard deviation for the parameter $\beta$ of the linear model, excluding countries with a GDP per capita below a given threshold in units of thousand dollars, from the {\it extended} set of 125 countries.
 \label{plotbetaGDP}}
\end{center}
\end{figure}

\begin{figure}
\begin{center}
\vspace*{3mm}
\includegraphics[scale=0.65]{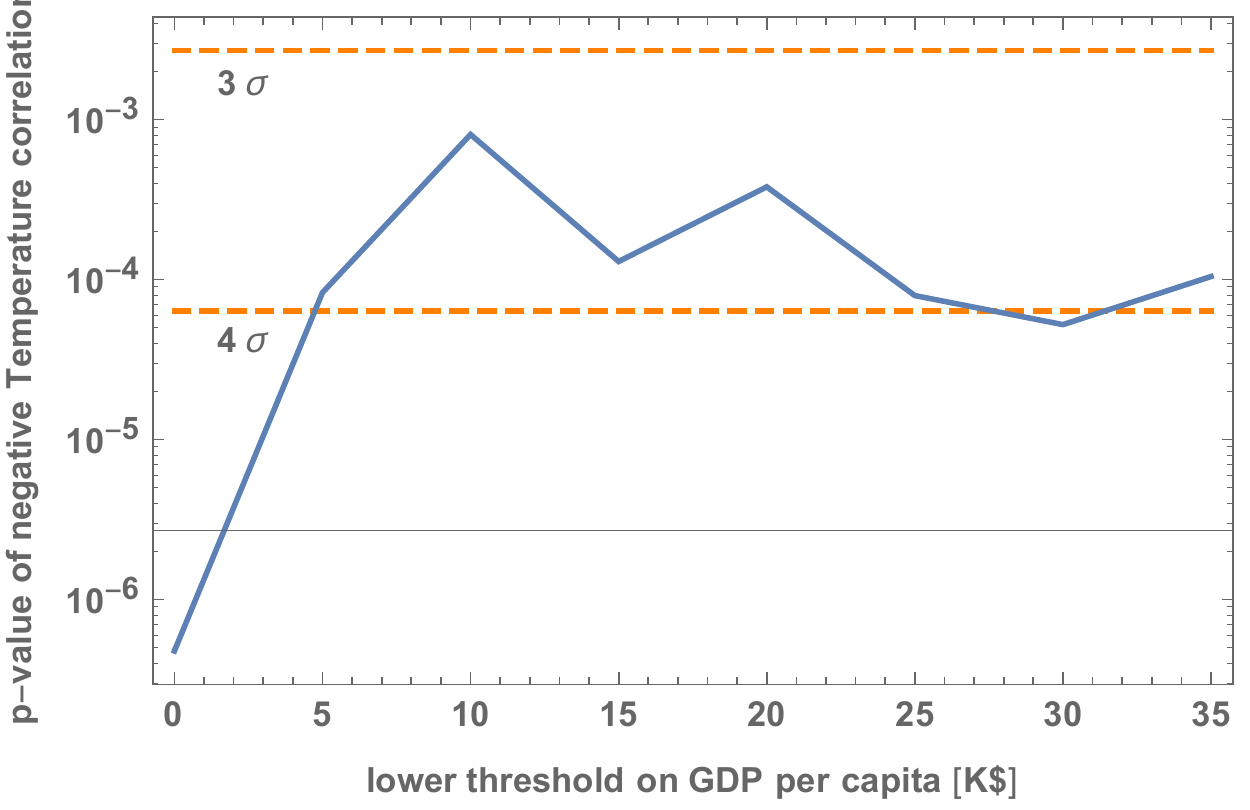} 
\caption{We show the significance ($p$-value)  for a nonzero parameter $\beta$, excluding countries with a GDP per capita below a given threshold in units of thousand dollars, from the {\it extended} set of 125 countries.
 \label{plotpvalue}}
\end{center}
\end{figure}

Finally we have tested the existence of a possible bias on the data: the fact that poor countries have less intense testing. This could in principle be a source of major bias, since many countries with low income are located in warm regions. In order to discard such a bias we have analyzed the existence of a nonzero linear correlation $\beta$ on subsamples of the {\it extended} dataset, by excluding countries with low income.
More specifically we have set a  threshold on the GDP per capita~\footnote{We used here data from https://ourworldindata.org/ on real GDP per capita, for the year 2017.},  and checked whether the correlation is still there, excluding countries below such a threshold from the analysis. We show in Fig.~\ref{plotbetaGDP} our results: we find a correlation to exist, rather independently on the threshold that we applied. The significance of a nonzero beta ($p$-value) is plotted in Fig.~\ref{plotpvalue} and remains always between $5\cdot 10^{-7}$ and $8\cdot 10^{-4}$.

In addition, we have also checked for a correlation between the growth rate $\alpha$ and the GDP per capita, shortly $GDP$. We find {\it no} significant correlation in the {\it base} and {\it intermediate} datasets, while we find a negative correlation in the {\it extended} dataset, with $p$-value $=0.0012$. This is not so surprising, since the {\it extended} dataset contains many low-income countries, where the disease has arrived later, and where most likely testing is not intense enough. For this dataset we performed thus a linear fit with two variables, $GDP$ and $T$. Results are shown in Table~\ref{tabCOV}. The dependence on $T$ is still highly significant, with $p$-value $\simeq  0.000048$ and the best-estimate is $\beta\simeq-0.0031$. As expected, $T$ also has non-negligible correlation with the GDP per capita.

\begin{table}[!htb]

\begin{tabular}{l|llll}
 \text{} & \text{Estimate} & \text{Standard Error} & \text{t-Statistic} & \text{p-Value}
   \\
\hline
 1 & 0.2186 & 0.01795 & 12.17 & $8.15^{-23}$ \\
 \text{GDP} & $6.165 \cdot 10^{-7}$ &
   $3.78\cdot10^{-7}$ & 1.627 & 0.1061 \\
 T & -0.003118 & 0.0007397 & -4.215 & 0.000048 \\
\end{tabular}

\vspace*{1em}
    \begin{minipage}{.5\linewidth}
\begin{tabular}{|l|c|c|c|c|}
 \hline
$R^2$ &  $0.2$ \\ \hline
$T-GDP$ correlation & $0.41$     \\  \hline
 \end{tabular}
    \end{minipage} 

 \caption{In the top panel: best-estimate, standard error ($\sigma$), $t-$statistic and $p-$value for the parameters of the linear interpolation in two-variables, temperature (T) and GDP per capita ($GDP$), for the {\it extended} set of 125 countries. In the bottom panel: $R^2$ and correlation coefficient ({\it i.e.} normalized off-diagonal element of the covariance matrix) between $T$ and $GDP$.}
 \label{tabCOV}
\end{table}

\section{Discussion and Conclusions} \label{conclusions}

We have collected data for countries that had at least 12 days of data after a starting point, which we fixed to be at the threshold of 30 confirmed cases. We considered three datates: a {\it base} dataset with 42 countries, collected on March 26th, an {\it intermediate} dataset with a total of 88 countries, collected on April 1st, and an {\it extended} dataset with a total of 125 countries, collected on April 14th. We have fit the data for each country with an exponential and extracted the exponents $\alpha$, for each country. Then we have analyzed such exponents as a function of the temperature $T$, using the average temperature for the month of March (or slightly earlier in some cases), for each of the selected countries.

For the {\it base} dataset we have shown that  the growth rate of the transmission of the COVID-19 has a decreasing trend, as a function of $T$, at $99.66\%$ C.L. ($p$-value 0.0034). In this fit $R^2=0.196$. In addition, using a quadratic fit, we have shown that a peak of maximal transmission seems to be present in this dataset at around $(7.7\pm 3.6)^\circ C$. Such findings are in good agreement with a similar study, performed for Chinese cities~\cite{Wang2020.02.22.20025791}, which also finds the existence of an analogous peak and an overall decreasing trend. Other similar recent studies~\cite{Araujo2020.03.12.20034728,Bukhari2020,Wang2020,Sajadi2020} find results which seem to be also in qualitative agreement.

For the {\it intermediate} dataset we also found a decreasing slope $\beta$. This is smaller in absolute value, but the significance remains high, since a zero slope is excluded at $99.86\%$ C.L. ($p$-value 0.0014). For this fit we found $R^2=0.11$.

Finally for the {\it extended } dataset we found a very highly significance for a negative $\beta$, $p$-value $5\cdot 10^{-6}\sim 5\cdot 10^{-7}$ (depending on the treatment of some anomalous cases), which would translate in a $4.5\sigma \sim5\sigma$ detection, in the language of Gaussian distributions. Here $R^2=0.16\sim 0.2$.

For all datasets we also tested the influence of a possible large bias: the fact that poorer countries have less intense testing, which might be in principle partially degenerate with effects of temperature. Our analysis indicate that this should not be a major issue: by excluding countries with low income from the analysis we find small variations on the best-fit value of $\beta$, and the significance of the correlation $\beta$ remains very high, with $p$-value $8\cdot 10^{-4}$ or less. We have also checked for a correlation between the GDP per capita and $\alpha$: we find  a significant correlation only in the {\it extended} dataset. This should be probably interpreted as the fact that poorer countries do not  have enough testing capabilities. However, after taking into account of this variable, the dependence on $T$ remains highly significant.

The decrease at high temperatures is expected, since the same happens also for other coronaviruses~\cite{Chan:2011}. 
It is unclear instead how to interpret the decrease at low temperature (less than $8^\circ C$), present in the {\it base} dataset. This could be a statistical fluctuation, since it is not present in the {\it intermediate}  and {\it extended} datasets. One possible reason for this decrease, if real, could be the lower degree of interaction among people in countries with very low temperatures, which could slow down the propagation of the virus.

A general observation is also that a  large scatter in the residual data is present, clearly due to many other systematic factors, such as variations in the methods and resources used for collecting data and variations in the amount of social interactions, due to cultural reasons. Further study is required to assess the existence and the relevance of such factors.

As a final remark, our findings can be very useful for policy makers, since they support the expectation that with growing temperatures the coronavirus crisis should become milder in the coming few months, for countries in the Northern Hemisphere. As an example the estimated doubling time, with the quadratic fit, at the peak temperature of $7.7^\circ C$ is of 2.6 days, while at $26^\circ C$ is expected to go to about 4.6 days. The linear fit implies an increase in the doubling time by $50\%$ (or $40\%$), going from $5^\circ$ C to $25^\circ$ C., using the estimate from the {\it extended} dataset (or the {\it extended} dataset, taking into account of the GDP per capita, at a reference value of 40 thousand dollars).   For countries with seasonal variations in the Southern Hemisphere, instead, this should give motivation to implement strong lockdown policies before the arrival of the cold season.

We stress that, in general, it is important to fully stop the propagation, using strong lockdown, testing and tracking policies, taking also advantage of the warmer season, and before the arrival of the next cold season.

\begin{acknowledgments}
We would like to acknowledge Viviana Acquaviva, Alberto Belloni, \'Angel J. G\'omez Pel\'aez, Jordi Miralda and Giorgio Torrieri,  for useful discussions and comments.
\end{acknowledgments}

\bibliographystyle{unsrt}
\bibliography{COVID}

\end{document}